\documentclass[aps,prd,twocolumn,nofootinbib,floatfix]{revtex4-1}
\usepackage{setspace}
\usepackage{amsthm}
\theoremstyle{definition}

\usepackage{graphicx}
\usepackage{dcolumn}
\usepackage{multirow}
\usepackage{subfigure}
\usepackage{times,mathptm}
\usepackage{float}
\usepackage{color}
\usepackage{amsmath,amsfonts}
\usepackage{mathptmx}
\usepackage{mathrsfs}
\usepackage{bbm}
\usepackage{bm}
\usepackage{xfrac}
\usepackage{url}

\newcommand{\beq}{\begin{equation}}
\newcommand{\eeq}{\end{equation}} 
\newcommand{\bea}{\begin{eqnarray}}
\newcommand{\eea}{\end{eqnarray}}

\newcommand{\E}{\mathcal{E}}

\renewcommand{\d}{\delta}

\renewcommand{\l}{\lambda}

\newcommand{\T}{{\cal T}}

\renewcommand{\b}{\beta}
\renewcommand{\a}{\alpha}

\newcommand{\tr}{\text{Tr}}
\newcommand{\hk}{\hat{k}}

\newcommand{\vx}{{\vec{x}}}
\newcommand{\vy}{{\vec{y}}}

\newcommand{\n}{\nu}
\newcommand{\m}{\mu}
\newcommand{\q}{\psi}
\newcommand{\qbar}{\overline{\psi}}
\newcommand{\g}{\gamma}

\renewcommand{\th}{\theta}

\newcommand{\dg}{\dagger}
\newcommand{\non}{\nonumber}

\newcommand{\rf}[1]{(\ref{#1})}
\newcommand{\ra}{\rightarrow}

\renewcommand{\vec}[1]{\bm #1}

\usepackage{ulem}

\begin{document}

\title{Excitations of isolated static charges in the charge $q=2$ abelian Higgs model}

\bigskip
\bigskip

\author{Kazue Matsuyama}
\affiliation{Physics and Astronomy Department \\ San Francisco State
University  \\ San Francisco, CA~94132, USA}
\bigskip
\date{\today}
\vspace{60pt}

\singlespacing
\begin{abstract}
  We present lattice Monte Carlo evidence of stable excitations of isolated static charges in the Higgs phase of the charge $q=2$ abelian Higgs model. These localized excitations are excited states of the interacting fields surrounding the static charges. Since the $q=2$ abelian Higgs model is a relativistic version of the Landau-Ginzburg effective action of a superconductor, we conjecture that excited states of this kind might be relevant in a condensed matter context.  Taken together with recent related work in SU(3) gauge Higgs theory, our result suggests that a massive fermion excitation spectrum may be a general feature of gauge Higgs theories.
 \end{abstract}
%
%
\maketitle
 
\singlespacing

\section{Introduction}

    Physical states in gauge field theories are gauge invariant, and this property implies that a static charge is necessarily accompanied by a surrounding field.\footnote{In the electroweak theory, the identification of gauge invariant operators creating particles in the asymptotic spectrum goes back to t Hooft \cite{tHooft:1979yoe} and Frohlich et al.\ \cite{Frohlich:1981yi} (see also Maas \cite{Maas:2019nso}).}  This could be a Coulomb field extending to infinity, as in free field electrodynamics,  or the charge of the state could be neutralized in some way by other charged dynamical fields.    In an interacting theory in which the surrounding field interacts with itself, there could in principle be a spectrum of localized quantum excitations of the surrounding field.  This is certainly true for a static quark-antiquark pair in the confining phase of a pure gauge theory.  In that case the color electric field associated with the pair of color charges is collimated into a flux tube,
and that flux tube can exist in a number of vibrational modes, as has been shown in various lattice Monte Carlo simulations \cite{Juge:2002br,Brandt:2018fft}.  By contrast, in free electrodynamics, any disturbance of the field surrounding a static charge can be viewed as the creation of some set of photons superimposed on a Coulombic background.  In that case there are no stable (or metastable) localized excitations.  What has not been studied in much detail is whether such excitations can exist in non-confining, but still interacting, gauge Higgs theories.

   Recently Greensite \cite{Greensite:2020lmh} has shown that there is indeed a spectrum of localized excitations
around an isolated fermion in SU(3) gauge Higgs theory, in the Higgs phase of the theory in four spacetime dimensions.
This raises the question of whether such an excitation spectrum is a general feature of gauge Higgs theories, particularly
those of physical interest such as effective theories of superconductivity, and the electroweak sector of the Standard Model.
Non-perturbative studies in the electroweak theory are complicated by the chiral nature of the gauge theory.  So we focus here
on a simple abelian gauge Higgs theory, namely the charge $q=2$ abelian Higgs model, which is a relativistic generalization
of the Landau-Ginzburg effective model of superconductivity.   In this article we will show that
stable localized excitations of the massive photon and Higgs fields surrounding a static charge can in fact exist in this theory,
at least in some regions of the phase diagram.  We believe this finding may be relevant to condensed matter systems, although our present work is limited to this result in the relativistic model.  Application of these methods to a more realistic model of superconductivity, and to the chiral gauge theories of interest to particle physics, is a topic which we defer to later work.
 
   Our strategy is to compute, via lattice Monte Carlo simulations, the energy (above the vacuum) of the ground state containing two static sources of opposite charge, and the energy of a certain excited state of this charge pair whose construction we describe. 
If the difference in energies is less than the photon mass, then the excited state is stable. This is what we will show below.

\section{Fermion excitation spectrum}

Our starting point is the lattice action of the abelian Higgs model  
\bea
     S  &=& - \beta \sum_{plaq}  \mbox{Re}[U_\m(x)U_\n(x+\hat{\m})U_\m^*(x+\hat{\n}) U^*_\n(x)]  \non \\
         & &       - \gamma \sum_{x,\m}  \mbox{Re}[\phi^*(x)U^2_\m(x) \phi(x+\widehat{\m})] \ . 
\label{Sgh}
\eea
Here the scalar field has charge $q=2$ (as do Cooper pairs), and 
for simplicity we impose a unimodular constraint,  ${\phi^*(x) \phi(x) = 1}$, corresponding to the
${\lambda \ra \infty}$ limit of a Mexican hat potential $\lambda (\phi \phi^* - \gamma)^2$, followed by a rescaling to $|\phi|=1$.
We then consider physical states containing a static fermion and anti-fermion at sites $\vx,\vy$, each of $\pm 2$ units of electric
charge, of the form  
\beq
         |\Phi_\a(R)\rangle = Q_\a(R) |\Psi_0\rangle \ ,
\label{Phistates}
\eeq
where $\Psi_0$ is the vacuum state and
\beq
         Q_\a(R) = [\qbar(\vx) \zeta_\a(\vx)] ~\times~  [\zeta^*_\a(\vy) \q(\vy)] \ .
  \label{Qop}
\eeq
Here the $\qbar,\q$ are operators creating double-charged static fermions of opposite charge, transforming as ${\q(x) \ra e^{2i\th(x)} \q(x)}$, and the $\{\zeta_\a(x)\}$  are a set of operators,
which may depend on some (possibly non-local) combination of the Higgs and gauge fields, also transforming as ${\zeta(x) \ra e^{2i\th(x)} \zeta(x)}$, under a gauge transformation
$U_\m(x) \ra  \exp(i\th(x)) U_\m(x) \exp(-\th(x+\hat{\mu}))$.  One possible choice for $\zeta$ is the Higgs field $\phi(x)$.  Another set is provided by eigenstates $\zeta = \xi_\a$ of the covariant Laplacian, where
 \beq
   (-D_iD_i)_{xy}\xi_\a({\vy};U) = \l_\a\xi_\a({\vx};U)
  \label{lap}
  \eeq 
and
 \beq
   (-D_iD_i)_{xy} = \sum_{k=1}^3 [2\delta_{\vx\vy} - U^2_k({\vx}) \delta_{{\vy, \vx} + \hk} - U^{* 2}_k ({\vx} - \hk)\delta_{{\vy, \vx} - \hk}] \ .
  \label{dd}
 \eeq 
 Because the covariant Laplacian depends only on the squared link variable, the $\xi_\a(x;U)$, which we have elsewhere referred to as  ``pseudomatter''  fields \cite{greensite2017},  transform like $q=2$ charged matter fields, with the one difference that, unlike matter fields, they do not transform under a global transformation in the center of the gauge group (which for U(1) is simply the group itself).  Pseudomatter fields depend nonlocally on the gauge fields, and the low-lying eigenstates and eigenvalues of the covariant Laplacian, which is a sparse matrix, can be computed numerically via the Arnoldi algorithm \cite{arpack}.\footnote{Eigenstates of the lattice covariant Laplacian were originally introduced by Vink and Wiese \cite{vink1992} to define a variant of Landau gauge which would be free from Gribov ambiguities.}  In our calculation we make use of the four lowest-lying Laplacian eigenstates and the Higgs field to construct the $\Phi_\a$,
 defining
 \beq
  \zeta_i(x) =  \left\{ \begin{array}{cl} 
                     \xi_i(x) &  i=1,2,3,4 \cr
                     \phi(x)  &  i=5 \end{array} \right. \ .  
\eeq
In general the five states $\Phi_\a(R)$ are non-orthogonal at finite $R$.  Of course $\phi(x)$ is a $q=2$ matter field, rather than pseudomatter field.
 
We express the operator $Q_\a$ in eq.\ \rf{Qop} in terms of a non-local operator $V_\a(\vx,\vy;U)$ 
\bea
 Q_\a(R) &=& \qbar(\vx) V_\a(\vx,\vy;U) \q(\vy) \non \\
V_\a(\vx,\vy;U) &=&   \zeta_\a(\vx;U) \zeta^*_\a(\vy;U)   \ ,
\eea 
and also define $\T = e^{-(H-\E_0)}$ as the Euclidean time evolution operator of the lattice abelian Higgs model.  This is the operator corresponding to the transfer matrix, multiplied by a constant $e^{\E_0}$ where $\E_0$ is the vacuum energy, evolving states for one unit of discretized time.  Let
 \bea
          [\T]_{\a\b} &=& \langle \Phi_\a | e^{-(H-\E_0)} | \Phi_\b \rangle  = \langle Q_\a^\dg(R,1) Q_\b(R,0) \rangle \non \\
          \left[ O \right]_{\a\b} &=&  \langle \Phi_\a | \Phi_\b \rangle  = \langle Q_\a^\dg(R,0) Q_\b(R,0) \rangle \non \\
\label{Tmn}
\eea 
denote matrix elements of $\T$, in the five non-orthogonal states $\Phi_\a$, with $[O]$ the matrix of overlaps
of such states.
We obtain the five orthogonal eigenstates of $\T$ in the subspace of Hilbert space spanned
by the $\Phi_\a$ by solving the generalized eigenvalue problem
 \beq
         [\T]_{\a\b} \upsilon_\b^{(n)} = \l_n [O]_{\a\b}\upsilon_\b^{(n)} \ ,
\label{general}
\eeq
with eigenstates denoted
\beq
          \Psi_n(R) = \sum_{\a=1}^3 \upsilon^{(n)}_\a \Phi_\a(R) \ .
\eeq 
and ordered such that $\l_n$ decreases with $n$.
We then consider evolving the states $\Psi_n$ in Euclidean time
\bea
         \T_{nn}(R,T) &=& \langle \Psi_n | e^{-(H-\E_0)T} | \Psi_n \rangle  \non \\
                                 &=& \upsilon^{*(n)}_\a \langle \Phi_\a | e^{-(H-\E_0)T}  | \Phi_\b \rangle \upsilon^{(n)}_\b   \non \\
                                 &=& \upsilon^{*(n)}_\a  \langle Q_\a^\dg(R,T) Q_\b(R,0) \rangle \upsilon^{(n)}_\b  \ ,
\label{TT}
\eea
where Latin indices indicate matrix elements with respect to the $\Psi_n$ rather than the $\Phi_\a$, and there is a sum
over repeated Greek indices.

To calculate this expression, we first define timelike $q=2$ Wilson lines of length $T$ 
\beq
          P(\vx,t,T) = U^2_0(\vx,t) U^2_0(\vx,t+1)...U^2_0(\vx, t +T-1) \ .
\eeq
After integrating out the massive fermions, whose worldlines lie along timelike Wilson lines, we have
\bea
       & & \langle Q_\a^\dg(R,T) Q_\b(R,0) \rangle \non \\
       & & = \langle \tr[V^\dg_\a(\vx,\vy;U(t+T)) P^\dg(\vx,t,T) V_\b(\vx,\vy;U(t)) P(\vy,t,T)]  \rangle \ . \non \\
\eea
On general grounds, $\T_{nn}(R,T)$ is a sum of exponentials
\bea
      \T_{nn}(R,T)  &=&   \langle \Psi_n(R) | e^{-(H-\E_0)T} | \Psi_n(R) \rangle \non \\
                            &=&   \sum_j |c^{(n)}_j(R)|^2 e^{-E_j(R) T} \ ,
\label{ERT}
\eea
where $c_j^{(n)}(R)$ is the overlap of state $\Psi_n(R)$ with the j-th energy eigenstate of the abelian Higgs theory containing a static fermion-antifermion pair at separation $R$, and $E_j(R)$ is the corresponding energy eigenvalue minus the vacuum energy.

    Of course one might expect that the $\T_{nn}(R,T)$ will all rapidly converge, in Euclidean time $T$, to a constant times
$\exp(-E_1 T)$, where $E_1$ is the ground state energy. This will be true for all $n$ {\it unless} one or more of the 
$|\Psi_n(R) \rangle$, constructed as just described, has only a very small overlap with the true ground state.  In that case the
exponential falloff may be dominated  by, e.g., the energy of the first excited states, at least for some moderate range of $T$.  In that situation it would be possible to extract the energy of that excited state in a simple way, without a multi-parameter fit to a sum of exponentials.

\section{Numerical results}      
      
\begin{figure}[htb]
 \includegraphics[scale=0.6]{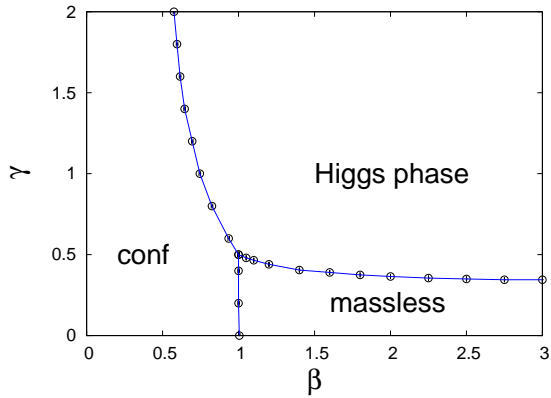}
 \caption{Phase diagram of the $q=2$ abelian Higgs model computed in \cite{matsuyama2019} ( ``conf''
 denotes the confinment phase).}
  \label{thermo}
 \end{figure}

   We proceed to the numerical results.  The phase diagram of the $q=2$ abelian Higgs model was first 
obtained from a lattice Monte Carlo simulation by Ranft et al in \cite{ranft1983}, and more recently and 
accurately by Greensite and the author in \cite{matsuyama2019}, with the result shown in Fig.\ \ref{thermo}.  
We are interested in determining $E_n(R)$ in the Higgs phase, and, because the calculation involves fitting 
exponential decay, we would like both the mass of the photon and the energies $E_n(R)$ to be not much larger 
than unity in lattice units.  For this reason we choose to work at the edge of the phase diagram shown in
Fig.\ \ref{thermo}, just above the massless-to-Higgs transition line at $\b=3, \g=0.5$.

 \begin{figure}[htb]
 \includegraphics[scale=0.6]{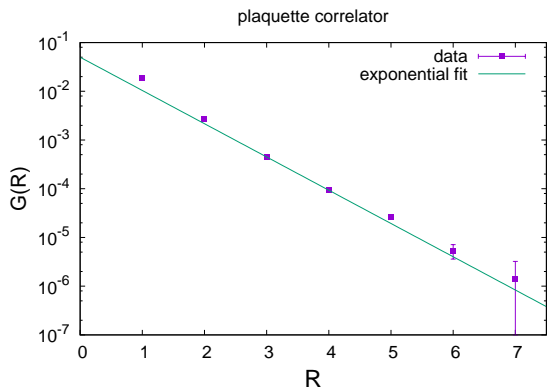}
 \caption{The plaquette-plaquette correlator computed for ${\b=3, \g=0.5}$. The photon mass is obtained from the slope of the line shown.}
  \label{photon}
 \end{figure}
    We compute the photon mass from the gauge invariant on-axis plaquette-plaquette correlator with the same
$\m \n$ orientation
 \bea
         G(R) &=& \bigg\langle {\rm Im}[U_\m(x)U_\n(x+\hat{\m})U_\m^*(x+\hat{\n}) U^*_\n(x)]  \non\\
     & & \times {\rm Im}[U_\m(y)U_\n(y+\hat{\m})U_\m^*(y+\hat{\n}) U^*_\n(y)] \bigg\rangle \ ,
\eea
where $y = x + R \hat{k}$, and $\hat{k}$ is a unit vector orthogonal to the $\hat{\m},\hat{\n}$ directions.  
The result for the $\b=3, \g=0.5$ parameters we have chosen is shown in Fig.\ \ref{photon}.  
From an exponential fit, disregarding the initial points, we find a photon mass of $m_\g = 1.57(1)$ in lattice units.  
Data was obtained on a $16^4$ lattice with 1,600,000 sweeps and data taken every 100 sweeps.  We have checked that if the
calculation is done just below the transition, in the massless phase, then $G(R)$ is fit quite well by a $1/R^4$ falloff, as expected. 
 
  \begin{figure}[t!]
 \includegraphics[scale=0.6]{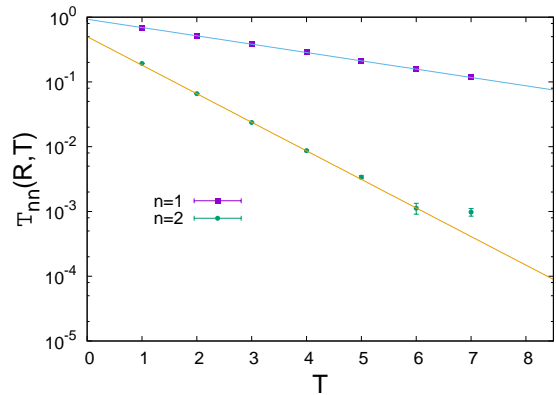}
 \caption{$\T_{nn}(R,T)$ vs.\ $T$ for $n=1,2$ at fixed $R=6.93$ on a $16^4$ lattice}
  \label{fit12}
 \end{figure}
 
 \begin{figure}[htb]
 \includegraphics[scale=0.6]{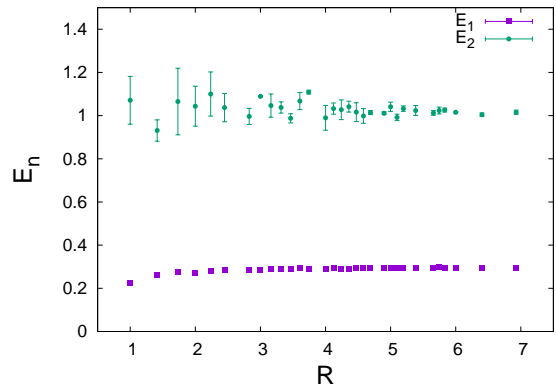}
 \caption{Energy expectation values $E_n(R)$ vs.\ $R$ for $n=1$ and $n=2$, obtained from a fit to a single
 exponential. }
  \label{ER12}
 \end{figure}
 
The energies $E_n(R)$ for $n=1,2$ are also obtained by fitting the data for $\T_{nn}(R,T)$ vs.\ $T$, at each $R$, to an exponential falloff.  An example of these fits at $R=6.93$ is shown is Fig.\ \ref{fit12}.   The data and errors were obtained from ten independent runs, each of 77,000 sweeps after thermalization, with data taken every 100 sweeps,
computing $\T_{nn}$ from each independent run. The lattice volume was again $16^4$, with couplings $\b=3, \g=0.5$. The points shown are the average of the ten sets, with the error taken as the standard error of the mean. The fits shown in Fig.\ \ref{fit12} are through the points at $T=2-5$,  with $E_1=0.2929(6)$ and 
$E_2(R) = 1.01(1)$ in this case.  The results of fits of this type, at all $R$, are displayed in Fig.\ \ref{ER12}.  

\begin{figure}[h!]
 \includegraphics[scale=0.6]{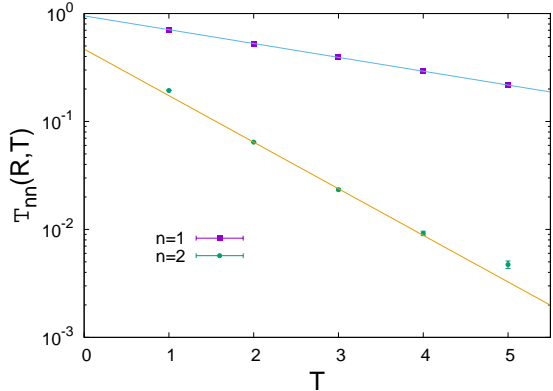}
 \caption{Same as Fig.\ \ref{fit12}, but on a $12^4$ lattice.}
  \label{fit12v12}
 \end{figure}

We note that the last data point in Fig.\ \ref{fit12}, at $T=7$, lies above the straight line on a log plot.  This is systematic, it is found at all $R$, and the question is whether it is a finite size effect.  To check this we can make the same computation, with the same number of sweeps,
only on a $12^4$ lattice.  The corresponding result at $R=6.93$ is shown in Fig.\ \ref{fit12v12}.  This time a fit through the points $T=2-4$ yields $E_2(R)=0.99(2)$, consistent with the larger volume result. Here we see that the last data
point, this time at $R=5$, also lies a little above the straight line fit, and again this effect is seen at all $R$.  This fact indicates that the deviation of the last data point from the fit to the other points is probably a finite size effect.

    The fact that $\T_{11}(R)$ is fit by a single exponential, a straight line on a log plot, starting at $T=1$, was certainly not obvious from the start.  It implies that $\Psi_1(R)$ must be very close to the ground state, rather than evolving to the ground state in
Euclidean time.  Convergence of $\Psi_2$ to a single exponential fit is also rapid, and is achieved after two time steps.  We reserve a discussion of fitting details to an Appendix.
    
 \begin{figure}[t!]
 \includegraphics[scale=0.6]{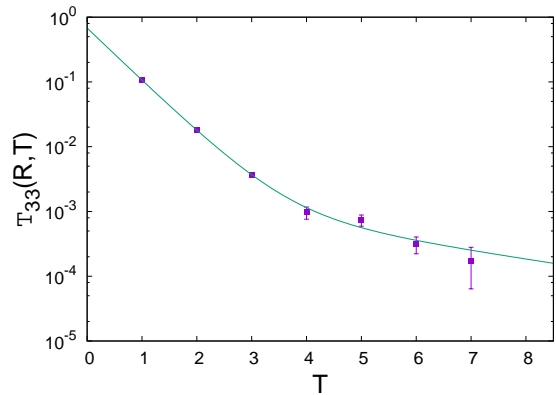}
 \caption{$\T_{33}(R,T)$ vs.\ $T$ at fixed $R=6.93$.  The fit shown is to the
 sum of exponentials in eq.\  \rf{T33fit}.}
  \label{fit3}
 \end{figure}   

\begin{figure}[htb]
 \includegraphics[scale=0.6]{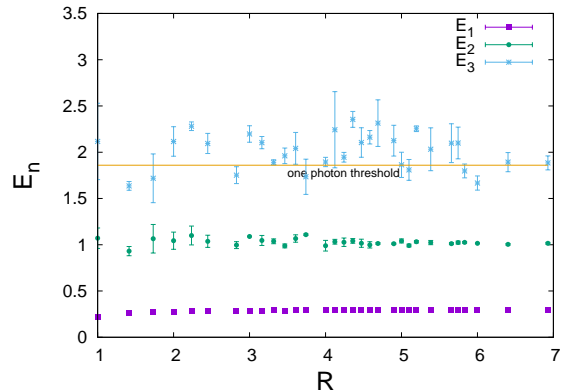}
 \caption{Energy expectation values $E_n(R)$ vs.\ $R$ for $n=1, 2, 3$ together with the one photon threshold.  The energy
 $E_3$ is obtained from a fit to two exponentials, as explained in the text.}
  \label{ER123}
 \end{figure}

    The data for $\T_{33}(R,T)$ simply does not fit a single exponential at any $R$, for the range of $T$ at our disposal.  To try and extract some information nonetheless, we can try to fit the data to a sum of three exponentials
\beq
          \T_{33}(R,T)  \approx a_1(R) e^{-E_1 T} + a_2(R) e^{-E_1 T} + a_3(R)  e^{-E_3 T} \ ,
\label{T33fit}
\eeq
where $E_1=0.29, E_2=1.02$ are taken from the previous fits.  A sample fit, again at $R=6.93$, is shown in Fig.\ \ref{fit3}.
Obviously one cannot be very impressed by a four parameter fit through a handful of data points.  What's more, there is no compelling reason to stop at three exponentials.\footnote{It is worth noting, however, that $a_1(R)$ is three orders of magnitude smaller than $a_2(R),a_3(R)$, indicating that $\Psi_3$ is almost orthogonal to the true ground state.}   But we do what we can; the idea here is to see if there is any indication of a second stable excited state, although the numerical value for $E_3$ should be regarded with appropriate caution.   With that caveat in mind, the values of $E_1, E_2, E_3$, together with the one photon threshold, are displayed in Fig.\ \ref{ER123}.

    The one photon threshold is simply $E_1 + m_{photon} = 0.29 + 1.57(1) = 1.86(1)$ in lattice units.  The important observation is
that $E_2(R)$ lies well below this threshold, which implies that the first excited state of the static fermion-antifermion pair is \emph{stable}.  The second point to note is that $E_3(R)$ seems to lie above or near the one photon threshold.  The indications are that there is no second stable excited state.  States above the first excited state most likely lie above the threshold, and are probably combinations of the ground state plus a massive photon.

\begin{figure}[htb]
 \includegraphics[scale=0.6]{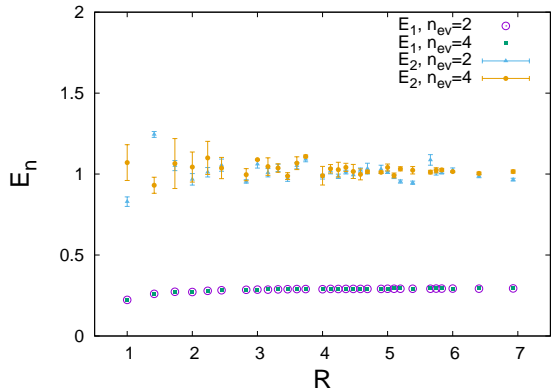}
 \caption{Comparison of $E_1,E_2$ obtained using $n_{ev}=2$ and $n_{ev}=4$ Laplacian eigenstates.}
  \label{compare}
 \end{figure}

     All of our results have been obtained using four pseudo matter fields, namely the four lowest lying eigenstates of the $q=2$ covariant Laplacian operator, and it is reasonable to ask what would be the result of changing this number.  Fig.\ \ref{compare} is a comparison of $E_1$ and $E_2$ values obtained from using $n_{ev}=2$ Laplacian eigenstates, with the values obtained using $n_{ev}=4$ Laplacian eigenstates.   
As can be seen in the figure, there is not much difference in the $E_2$ values, at least for $R \ge 3$, and the $E_1$ values cannot even be distinguished in the plot.

\section{Conclusions}

   To summarize, we have presented lattice Monte Carlo evidence for  the existence of a stable excitation of the quantized fields surrounding isolated static charges, in the Higgs phase of the $q=2$ abelian Higgs model in $D=4$ spacetime dimensions.  The $q=2$ abelian Higgs model is a close relative of the non-relativistic Ginzburg-Landau effective action of superconductivity.  So the obvious next question is whether excitations of the type seen in the abelian Higgs model would also be found in non-relativistic models of that kind.  If such excitations are found to exist in a realistic effective model, then the follow-up question is how they might be observed experimentally.   A further question is whether heavy fermions (or even light fermions) have a spectrum of excitations in the electroweak sector of the Standard Model.  Although the lattice regularization of chiral gauge theories is known to be problematic, perhaps something can still be done numerically using non-dynamical static charged sources.  We leave these possibilities for future investigation.

\acknowledgements{I would like thank Jeff Greensite for calling my attention to his recent work in SU(3) gauge Higgs theory, and for many helpful conversations.}
 
\appendix*
\section{Some fitting details}

     We begin by noting that our method involves solving the generalized eigenvalue equation \rf{general}, and an exact solution will provide eigenstates satisfying the orthogonality condition $\langle \Psi_i | \Psi_j \rangle = \d_{ij}$.  Surprisingly, the numerical solution of this eigensystem, by the standard Matlab {\tt \bf eig} routine (ultimately derived from LAPACK), shows a small $O(10^{-3})$ but non-negligible deviation from this orthogonality condition.  We have therefore made $\Psi_2$ orthogonal to $\Psi_1$ by subtracting its projection onto $\Psi_1$ (i.e.\ the first step of a Gram-Schmidt procedure).  This makes a small, but nonetheless noticeable, improvement in the single exponential fits to $\T_{22}$.
  
\begin{figure}[htbp]
 \includegraphics[scale=0.55]{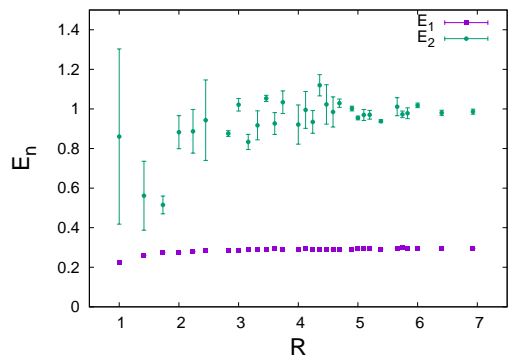}
 \caption{Same as Fig.\ \ref{ER12}, except $E_2$ is obtained from fits to data points in the range $T=3-6$, rather than $T=2-5$.}
 \label{ER12a}
 \end{figure}

\begin{figure}[htbp]
 \includegraphics[scale=0.55]{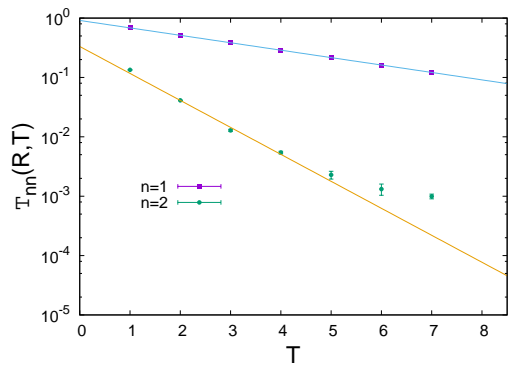}
 \caption{$T_{22}(R)$ vs.\ $T$ at $R=3.16$.  The reason for a discrepancy between
 a fit of data points at $T=2-5$, and $T=3-6$ is apparent.}
  \label{fitn10}
 \end{figure}

      We also note that at the larger $R>3$ values, on a $16^4$ lattice, the next-to-last data point at $T=6$ lies  mostly on or near the best exponential fit through the points at $T=2-5$. But this is not always the case, especially for lower $R$, and as a result a fit for data points in the range $T=3-6$, rather than $T=2-5$ often results in a high $\chi^2$.  We display in Fig.\ \ref{ER12a} the values of $E_2$ obtained from a fit in the $T=3-6$ interval.  In general the $E_2$ values cluster around $E_2=1$, as in the previous fit.  But there are large error bars for some of the points, especially at the lower $R$ values, and significant deviations from $E_2\approx 1$. Data points up to $R=2.5$ are obtained from fits with rather large $\chi^2$ values, and can be discarded simply on those grounds.  In Fig.\ \ref{fitn10}, corresponding to $R=3.16$, one can see the reason for these deviations: both the last data points for $\T_{22}$ at $T=7$ {\it and} the next-to-last data point at $T=6$ deviate very significantly from the fit in the $T=2-5$ range.  We are inclined to attribute both deviations to finite size effects, which seem especially apparent at lower $R$.      

\begin{figure}[htb]
 \includegraphics[scale=0.6]{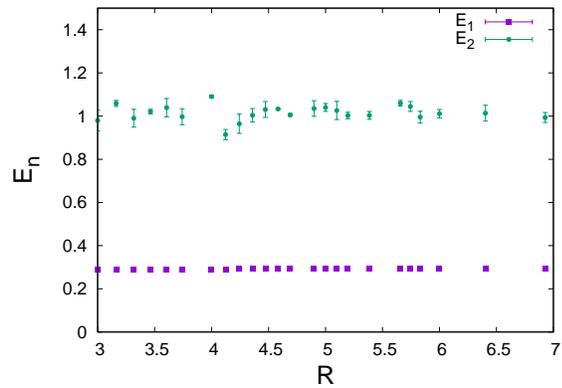}
 \caption{Energy expectation values $E_1, E_2$ vs.\ $R$  at $R>3$, obtained on a $12^4$ lattice.}
  \label{ER12v12}
 \end{figure}

     We conclude with a display, in Fig.\ \ref{ER12v12}, of $E_1,E_2$ obtained on a $12^4$ lattice volume.  As in the larger volume, the data for $E_2$ clusters around $E_2 \approx 1$, albeit with a few outliers.  These values, however, are obtained from a fit through only three data points at $T=2,3,4$, and also the $\chi^2$ values  of these fits tend to be significantly larger than unity, indicating a possible underestimate of the error bars.
 
\bibliography{U14d.bib}

\end{document}